\documentclass[aps,prd,superscriptaddress,twocolumn,showpacs]{revtex4-2}

\usepackage{graphicx}
\usepackage{ulem}
\usepackage{amssymb}

\usepackage{graphicx}
\usepackage{amsmath}
\usepackage{amssymb}
\usepackage{sistyle}
\usepackage{color}

\begin{document}

\title{Landau level collapse in graphene in the presence of \\ in-plane radial electric and perpendicular magnetic fields}

\author{I.O.~Nimyi}
\affiliation{Kyiv Academic University, 03142 Kyiv, Ukraine}
\affiliation{Institute for Theoretical Solid State Physics, IFW Dresden and
W\"{u}rzburg-Dresden
Cluster of Excellence ct.qmat, Helmholtzstr. 20, 01069 Dresden, Germany}

\author{V.~K\"{o}nye}
\affiliation{Institute for Theoretical Solid State Physics, IFW Dresden and
W\"{u}rzburg-Dresden
Cluster of Excellence ct.qmat, Helmholtzstr. 20, 01069 Dresden, Germany}

\author{S.G.~Sharapov}
\affiliation{Bogolyubov Institute for Theoretical Physics, National Academy of Science of Ukraine, 14-b Metrologichna Street, Kyiv, 03143, Ukraine}
\affiliation{Kyiv Academic University, 03142 Kyiv, Ukraine}

\author{V.P.~Gusynin}
\affiliation{Bogolyubov Institute for Theoretical Physics, National Academy of Science of Ukraine, 14-b Metrologichna Street, Kyiv, 03143, Ukraine}
\date{\today }

\begin{abstract}
It is known that in two-dimensional relativistic Dirac systems placed  in
orthogonal uniform magnetic and electric fields, the Landau levels collapse as the applied in-plane electric field reaches a critical value $\pm E_c$.
We study this phenomenon for a distinct field
configuration with in-plane constant radial electric field.
The Dirac equation for this configuration does not allow
analytical solutions in terms of known special functions.
The results are obtained by using both the WKB approximation and
the exact diagonalization and shooting methods.
It is shown that the collapse occurs for  positive values of
the total angular momentum quantum number,
the hole (electron)-like Landau levels collapse as the electric field
reaches the value $ +(-) E_c/2$.
The investigation of the Landau level collapse
in the case of gapped graphene shows a number of distinctive features
in comparison with the gapless case.

\end{abstract}



\maketitle

\section{Introduction}

It was Rabi \cite{Rabi1928ZP} who solved the just-discovered Dirac equation
in a homogeneous magnetic field in the symmetric gauge and showed
that the energies of the free electrons are quantized. This occurred
two years before the corresponding quantized levels were found in the nonrelativistic
quantum theory by Frenkel and Bronstein \cite{Frenkel1930} and Landau \cite{Landau1930}.
However the experimental exploration of the relativistic
Landau levels, in contrast to the nonrelativistic ones,
in condensed matter systems
became possible almost 80 years later
after the discovery of graphene \cite{Geim2005,Kim2005}.
Naturally, the most exciting are the properties
of the relativistic Landau levels that do not have their counterparts
for  standard electron systems and among them is the
Landau level collapse phenomenon predicted in Ref.~\cite{Lukose2007PRL} (see also Ref.~\cite{Peres2007JPCM})
and observed experimentally in Refs.~\cite{Singh2009PRB,Gu2011PRL}.

This phenomenon occurs when, in addition to a magnetic field $H$ applied perpendicular
to the sheet of graphene, an in-plane electric field $E$ is present.
It consists of the merging of the  Landau-level staircase when the applied
electric field reaches  a critical value $\pm E_c$ with $E_c = (v_F/c) H$ in
CGS units, where $v_F$ is the Fermi velocity.

There are several ways to understand the origin of the collapse.
The first one is based on the consideration of the motion
in  crossed electric and magnetic fields
\cite{Lifshitz1959UFN,Lifshitz1973book}, where
the motion  of  a quasiparticle having a dispersion $\mathcal{E}(\mathbf{p}) $
in crossed fields, can be viewed as a motion of a particle in the magnetic field
only with the modified dispersion law,
\begin{equation}
\mathcal{E}^\ast (\mathbf{p}) = \mathcal{E}(\mathbf{p}) - \mathbf{v}_0 \mathbf{p},
\end{equation}
with $\mathbf{v}_0 =  c \mathbf{E} \times \mathbf{H}/H^2$
being the drift velocity for the motion in the crossed fields.
Then the quasiclassical spectrum follows from the  Lifshitz-Onsager quantization condition,
\begin{equation}
\label{orbit-electric}
S(\mathcal{E}^\ast) = 2 \pi \hbar \frac{e H}{c} (n + \gamma_B ), \quad n=0,1,\ldots,
\end{equation}
where $S(\mathcal{E}^\ast)$ is the electron orbit area in the
momentum space, $\gamma_B$ is the
topological part of the Berry phase.
For the quadratic dispersion law $\mathcal{E} = p^2/(2 m)$ with $m$
being the effective mass and $p$ the absolute value of the momentum,
the area is
$S(\mathcal{E}^\ast)  = 2 \pi m \mathcal{E}^\ast $. Thus, one can see that
in this case the electric field does not change the distance between Landau levels.

The massive Dirac fermions with the dispersion,
$\mathcal{E} = \pm \sqrt{v_F^2 p^2 + \Delta^2}$, are characterized by the area
$S(\mathcal{E}) = \pi (\mathcal{E}^2 - \Delta^2)/v_F^2$, where $\Delta$ is the gap
in the quasiparticle spectrum.
The Lifshitz-Onsager quantization
condition Eqs.~(\ref{orbit-electric})  results in the spectrum in the crossed fields \cite{Alisultanov2014PB}
\begin{equation}
\label{LL-collapse}
\begin{split}
\mathcal{E}_n & = \mathcal{E}^\ast_n - \hbar k \frac{E}{H}, \\
\mathcal{E}^\ast_n & = \pm (1 - \beta^2 )^{3/4}\sqrt{\frac{ 2n \hbar v_F^2 e H}{c}
+ \frac{\Delta^2}{(1 - \beta^2 )^{1/2}}},
\end{split}
\end{equation}
where $k$ is the in-plane wave vector along the direction perpendicular to the electric field,
$\beta = v_0/v_F = c E/(v_F H)$,
the phase $\gamma_B=0$. For $\Delta =0$,
the spectrum Eqs.~(\ref{LL-collapse}) reduces the spectrum obtained by
an exact solution of the problem \cite{Lukose2007PRL,Peres2007JPCM}.
The Landau level collapse occurring at $|\beta| =1$ can be viewed as
a transition from the closed elliptic quasiparticle  orbits for
$|\beta| <1$ ($|v_0| < v_F$) to open hyperbolic orbits for $|\beta| >1$ ($|v_0| > v_F$)
\cite{Shytov2009SSC}.

Another elegant way to understand the origin of the collapse and even
to derive the spectrum Eqs.~(\ref{LL-collapse}) is described in Ref.~\cite{Lukose2007PRL}
(see also Refs.~\cite{Shytov2009SSC,Arjona20017PRB}).
One can employ the effective Lorentz covariance of the equation of
motion [see Eq.~(\ref{Dirac-E}] below), in which the graphene dispersion velocity $v_F$ plays the role of the speed of light $c$, and consider the corresponding Lorentz transformations for electromagnetic fields.

We assume that the magnetic  and electric fields are directed in the $z$
and $y$ directions, respectively. Then
under a ``boost'' in the $x$ direction with velocity $v$, these fields
transform as \cite{foot1}
\begin{subequations}
\label{Lorentz}
\begin{align}
\label{E}
E_y^\prime & = \gamma \left(E_y - \frac{v}{c} H_z \right),\\
\label{B}
H_z^\prime & = \gamma \left(H_z - \frac{v c}{v_F^2} E_y \right),
\end{align}
\end{subequations}
where $\gamma = 1/\sqrt{1 - v^2/v_F^2}$. When the velocity $v$ coincides
with the drift velocity $v_0 = c E/H$ in graphene,
the electric field disappears in the primed reference frame
and the magnetic field becomes
\begin{equation}
\label{B-prime}
H_z^\prime =\sqrt{1- \beta^2} H.
\end{equation}
Here we assumed that in the original frame $c |E| \leq v_F |H|$ (or $|\beta| \leq 1$).
Since the Dirac equation is Lorenz covariant,
the energies of Landau levels in the primed frame are known:
\begin{equation}
\label{spectrum-prime}
\mathcal{E}^\prime_n = \pm \sqrt{ 2n \hbar v_F^2 e H_z^\prime/c +\Delta^2}.
\end{equation}
The value $\Delta/v_F^2$ plays the role of
the mass in the Dirac theory and remains invariant under
Lorentz transformations. Considering that the energy is the zeroth component
of the energy-momentum vector and
doing the inverse boost transformation one recovers the spectrum
Eq.~(\ref{LL-collapse}).
The critical value $|\beta| =1$ corresponding to
the collapse of all levels $\mathcal{E}^\ast_n = \mathcal{E}^\prime_n \sqrt{1-\beta^2}$ is determined by the relationship
Eq.~(\ref{B-prime}) between $H_z^\prime$ and $H$.

\begin{figure}[!h]
\includegraphics[width=.4\textwidth]{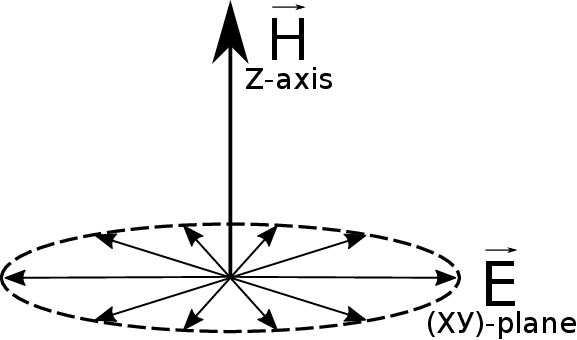}
\caption{Schematic figure for the electric and magnetic field configuration.
The radial electric field $E$ is in the plane of the graphene sheet and
the uniform magnetic field $H$ is perpendicular to this plane.}
\label{fig:1}
\end{figure}

The purpose of this paper is to study the Landau
levels and their collapse in another field configuration
with the magnetic field $H$ applied perpendicular to the infinite
graphene's plane and the in-plane constant radial electric field $E$
as shown in Fig.~\ref{fig:1}.
In practice, an approximately constant radial field can be created
inside a cylindric capacitor, where the electric potential
\begin{equation}
V(r) = V_0 \ln \frac{r}{a} \approx V_0 \left(\frac{r}{a}-1 \right),
\qquad b \leq r \leq a,
\end{equation}
with $a$ and $b$ being the external and internal radii, respectively.

Unlike the above-described  case of
orthogonal uniform magnetic and electric fields, the present
problem can not be solved exactly in a closed analytic form.
Thus, to investigate this problem, we employ
the semiclassical WKB method which leads to a transcendental equation
for the spectrum. These results are compared with the calculations
performed using the exact diagonalization and shooting methods.
It  turns out
that the WKB solutions are very close to the numerical ones for practically all
quantum numbers.

Following the above-mentioned arguments with the Lorentz transformation, one
may develop a qualitative understanding of the present problem.
However, one should keep in mind that because the electric field is radial
rather than unidirectional, the considered case rather resembles
an explanation of the origin of spin-orbit interaction in atomic physics.
Thus, the Thomas precession must be properly taken into
account (see, for example, Ref.~\cite{Jackson1999book}).
One can see from the transformation Eq.~(\ref{E}) that by
doing a boost to the coordinate system moving with the drift velocity
$v_0 = c E/H$, it is possible to remove the electric field locally.

We emphasize that due to Thomas precession
it is not sufficient to use the transformation Eq.~(\ref{B}), and Eq.~(\ref{B-prime})
has to be replaced by a more complicated unknown relationship
$H_z^\prime =\alpha(\beta) H$. The Landau-level collapse
would still be possible if an unknown function $\alpha(\beta)$ goes to zero
at some value of $\beta$. We later show that for the configuration of
the fields in Fig.~\ref{fig:1},
$\beta = -1/2$ for electrons and $\beta = 1/2$ for holes.

The paper is organized as follows.
In Sec.~\ref{sec:model}, the model for a single layer of graphene
with a magnetic field $H$ applied perpendicular to the layer
and an in-plane constant radial electric field is introduced.
Using the symmetry of the problem, it is reduced to
a system of radial equations. This system is
considered using the WKB method in Sec.~\ref{sec:WKB}.
In particular, the transcendental WKB equation for
the energy spectrum is derived in terms of complete elliptic integrals for
gapless graphene.
The technical details are
provided in  Appendices~\ref{sec:Appendix-int} and
\ref{sec:Appendix-int-beta=1/2}.
In the absence of an electric field, the WKB approximation
recovers the exact solution as discussed both in
Sec.~\ref{sec:E=0} and Appendix~\ref{sec:E=Delta=0}.
In Sec.~\ref{sec:results}, we obtain and discuss the energy spectra
obtained in the WKB  approximation and compare them with the
results of numerical computations performed using the exact
diagonalization and shooting methods.
The  gapped graphene case is considered  using numerical
methods in Sec.~\ref{sec:gapped}.
In the Conclusion  (Sec.~\ref{sec:conclusion}), we summarize the obtained results
and discuss their possible experimental observation.

\section{Model and main equations}
\label{sec:model}

We consider the stationary Dirac equation
\begin{equation}
\label{Dirac-E}
\left[-\hbar
v_{F}\left(\alpha_{1}iD_{x}+\alpha_{2}iD_{y}\right) + \Delta \alpha_{3} + V(\mathbf{r}) - \mathcal{E}  \right]
\Psi( \mathbf{r})=0,
\end{equation}
which describes  low-energy excitations in graphene (see, e.g., Ref.~\cite{Gusynin2007review}
for a review and notations)
and eigenenergy $\mathcal{E}$.
The $4\times 4$ $\alpha$-matrices $\alpha_{i}= \tau_3 \otimes \sigma_i$
and the Pauli matrices $\tau_i$, $\sigma_i$ (as well as the $2\times 2$ unit
matrices $\tau_0$, $\sigma_0$) act on the valley ($\mathbf{K}_\eta $ with
$\eta = \pm$)
and sublattice ($A,B$) indices, respectively, of the
four component spinors $\Psi^T = \left( \Psi_+^T, \Psi_-^T \right) =
\left( \psi_{AK_+}, \psi_{BK_+}, \psi_{BK_-}, \psi_{AK_-} \right)$.

We consider both the massless Dirac-Weyl fermions in the pristine graphene
and the massive Dirac fermions with the mass $\Delta$.
We recall that a global $A/B$ sublattice asymmetry gap
$2 \Delta \sim \SI{350}{K}$ can be introduced in graphene
\cite{Hunt2013Science,Gorbachev2014Science,Woods2014NatPhys,Chen2014NatCom}
when it is placed on top of hexagonal boron nitride (G/hBN) and
the crystallographic axes of graphene and hBN are aligned.

The  orbital effect of a perpendicular magnetic field $\mathbf{H} = \nabla \times \mathbf{A}$
is included via the covariant spatial
derivative  $D_j=\partial_j+(ie/\hbar c)A_j$ with $j=x,y$ and $-e<0$,
while the potential $V(\mathbf{r})$ corresponds to the
static electric field $e \mathbf{E} =\nabla V(\mathbf{r}) $.
The Zeeman interaction is neglected in this paper
(see, e.g., Ref.~\cite{Gusynin2007review}) and
the spin index is omitted in what follows.

We consider the  configuration of crossed magnetic and electric fields,
with the magnetic field applied perpendicular
to the infinite plane of graphene along the positive $z$ axis \cite{foot2} and the corresponding
vector potential is taken in the symmetric gauge
$(A_x,A_y)=(H/2)(-y,x)$ and
radial in-plane  electric field $E$ with the potential  $V(r)=eEr$ (see Fig.~\ref{fig:1}).

It is clear that the solution at the $\mathbf{K}_-$ point is obtained from the solution
at the $\mathbf{K}_+$ point by changing
$\Delta \to - \Delta$  and exchanging the spinor components
$\psi_A  \leftrightarrow \psi_B$, so in what follows we only consider
the $\mathbf{K}_+$ point.

Since the system has rotational symmetry,
it is natural to consider the problem in polar coordinates, where
\begin{equation}
\label{derivative-angle}
i D_{x} \pm D_{y}=e^{\mp i\phi}\left(i\frac{\partial}{\partial r}
\pm \frac{1}{r}\frac{\partial}{\partial\phi} \pm \frac{ie H r}{2\hbar c}\right).
\end{equation}
Accordingly, the total angular momentum
$J_z$ is conserved and we can represent $\Psi(\mathbf{r})$ in terms of the eigenfunctions of
$J_z=L_z+\sigma_z/2=-i\partial/\partial\phi+\sigma_z/2$ as follows:
\begin{equation}
\label{angular-spinor}
\Psi_+(\mathbf{r})=\left[\begin{array}{c}e^{i(j-1/2)\phi}f(r)\\ i e^{i(j+1/2)\phi}g(r)\end{array}\right],
\end{equation}
where $j=\pm 1/2,\pm 3/2,\ldots$ is the total angular momentum quantum number.
Then for the spinor
$\chi^T(r)=\left( f(r), g(r)\right)$,
we obtain the following system of equations written in a matrix form:
\begin{equation}
\label{Dirac-system-WKB}
\chi'(r)=\frac{1}{\hbar}N(r) \chi (r),
\end{equation}
where the prime denotes the derivative over $r$ and the matrix
\begin{equation}
\label{N}
N=\left[\begin{array}{cc}\frac{\hbar(j-1/2)}{r}+
\frac{e H r}{2c}&-\frac{{\cal E}+\Delta-V(r)}{v_{F}}\\ \frac{{\cal E}-\Delta-V(r)}{v_{F}}&-\frac{\hbar(j+1/2)}{r}-
\frac{e H r}{2c}\end{array}\right].
\end{equation}

For a constant electric field with $V(r)=eEr$,
one can rewrite the last equation in the form
\begin{equation}
\label{ABC-system}
\chi'(\rho)=N(\rho) \chi(\rho)=\left(\frac{A}{\rho}+B+C\rho\right)\chi(\rho),
\end{equation}
where
we introduced the dimensionless variable $\rho=r/l$ with
$l = \sqrt{\hbar c/e H}$ being the magnetic length; the prime now
denotes the derivative over $\rho$,
and the $2\times2$  $\rho$-independent matrices $A,B,C$ are, respectively,
\begin{equation}
\label{radial-system}
\begin{split}
A & =\left[\begin{array}{cc}j-1/2&0\\0&-j-1/2\end{array}\right], \\
B & =\left[\begin{array}{cc}0&
-\frac{l(\mathcal{E}+\Delta)}{\hbar v_F}\\ \frac{l(\mathcal{E}-\Delta)}{\hbar v_F}&0\end{array}\right],\\
C & =\left[\begin{array}{cc}1/2&\beta\\-\beta&-1/2\end{array}\right].
\end{split}
\end{equation}
The matrix $C$ contains the
important dimensionless parameter $\beta = c E/(v_F H)$
that describes the strength of the electric field relative to the
magnetic field. In this paper, we restrict ourselves to the $|\beta| \leq 1/2$ case.

Comparing the system of equations (\ref{ABC-system}) with
the corresponding system describing the Dirac fermions
in the uniform  magnetic
field and constant electric field in the $x$ direction
\cite{Lukose2007PRL,Peres2007JPCM} (see also Ref.~\cite{MacDonald1983PRB}),
one can see that the latter contains only two matrices $\tilde B + \tilde C x$.
The problem in the crossed uniform  fields in the Cartesian coordinates
is exactly solvable by diagonalizing the matrix $\tilde C$, while the present
problem with the radial electric field cannot  be solved analytically.
The situation is similar to the case of 2D Dirac fermions in a
constant magnetic field and Coulomb potential (see, e.g., Refs.~\cite{Khalilov2000PRA,Gamayun2011PRB,Zhang2012PRB,Sun2014PRB,Li2021}) and to the parabolic potential $V(r) \sim r^2$ \cite{Rodriguez-Nieva2016PRB},
which do not have analytical solutions in terms  of known special
functions.


Making in Eq.~(\ref{ABC-system}) the transformation of dependent variable
$\chi(\rho)=U(\rho) \xi(\rho)$, where $U(\rho)$ is a $2\times2$ matrix with ${\rm det}\, U(\rho) \neq0$,
we obtain a different system of equations,
\begin{equation}
\label{ABC-system-transform}
\xi'(\rho)=  M(\rho)\xi(\rho),
\end{equation}
with matrix $N(\rho)$  related to matrix $M(\rho)$  by the transformation
\begin{equation}
M(\rho)=U^{-1}(\rho) N(\rho)U (\rho)-U^{-1}(\rho) U'(\rho).
\end{equation}

Choosing the matrix $U$ proportional to the unit matrix $U= \sigma_0 e^{-a\rho^2/2}\rho^s$, one can see that matrix $B$
does not change, while  matrices $A$ and $C$ become
\begin{equation}
\label{matrices-ABC}
\begin{split}
A=\left[\begin{array}{cc}j-1/2-s&0\\0&-j-1/2-s\end{array}\right],\\ C=\left[\begin{array}{cc}1/2+a&\beta\\-\beta&-1/2+a\end{array}\right].
\end{split}
\end{equation}

Different sets of parameters $a$ and $s$ are appropriate for the consideration of the system (\ref{ABC-system-transform}),
for example, choosing $a=1/2$ and $s=-j-1/2$ one obtains the following
second-order differential equation for the lower component $\xi_2(\rho)$ of the spinor $\xi(\rho)$:
\begin{equation}
\label{2nd-order}
\xi_2'' -\left(\frac{2j}{\rho}+\rho + \frac{\beta}{\beta\rho-\epsilon+\delta}\right)\xi'_2+[(\beta\rho-\epsilon)^2-\delta^2]\xi_2=0.
\end{equation}
Here we introduced the dimensionless energy $\epsilon = l \mathcal{ E}/(\hbar v_F)$, and mass (gap) $\delta = l \Delta/(\hbar v_F)$.
This equation has three singular points, two regular at $\rho=0,\rho=(\epsilon-\delta)/\beta$  and an irregular one at infinity.
The singularity at $\rho=(\epsilon-\delta)/\beta$ is apparent [it is absent in the system (\ref{ABC-system-transform})].
The equation is similar to the confluent Heun equation but the singularity at infinity is more strong (with rank three according to the definition
of the rank of singular points of differential equations in the  Ref.~\cite{Slavyanov-book}).
The rank of the irregular singular point at infinity of the confluent Heun
equation is equal to two. Analytical results for such type of equations
are absent in the literature, thus it is
necessary to use another method to investigate the problem.

\section{WKB method}
\label{sec:WKB}

We employ the WKB method to study the spectrum
of bound states for the problem
described by Eqs.~(\ref{ABC-system-transform}) and (\ref{matrices-ABC}).
It is convenient to choose the exponent $s=-1/2$ and $a=0$ in Eqs.~(\ref{matrices-ABC}), so
the spinor $\xi(r) =\sqrt{r} \chi(r)$.
Restoring the dimensional variables,
we rewrite Eq.~(\ref{ABC-system-transform}) as follows:
\begin{equation}
\label{WKB-Dirac}
\xi'(r)=\frac{1}{\hbar}M\xi(r),
\end{equation}
where the matrix
\begin{equation}
\label{matrix-M}
\begin{split}
M & = \left[\begin{array}{cc}\frac{J}{r}+
\frac{e H r}{2c}&-\frac{\mathcal{ E}+\Delta-V(r)}{v_{F}}\\
\frac{\mathcal{ E}-\Delta-V(r)}{v_{F}}&-\frac{J}{r}-
\frac{e H r}{2c}\end{array}\right] \\
& = \frac{\hbar}{l}\left[\begin{array}{cc}\frac{j}{\rho}+\frac{\rho}{2}&\beta\rho-\epsilon-\delta\\
-(\beta\rho-\epsilon+\delta)& -\frac{j}{\rho}-\frac{\rho}{2} \end{array}\right].
\end{split}
\end{equation}
Here we denoted $J = \hbar j$ in the first line.
Eqs.~(\ref{WKB-Dirac}) and (\ref{matrix-M}) are similar
to the corresponding system written in Refs.~\cite{Khalilov2000PRA,Zhang2012PRB,Sun2014PRB,Rodriguez-Nieva2016PRB,Li2021}
(see Ref.~\cite{foot2}).

As one can see from the first line of Eq.~(\ref{matrix-M}), the system Eq.~(\ref{WKB-Dirac}) contains a small parameter $\hbar$ and it is possible to use
the standard scheme for solving asymptotically systems of linear differential equations \cite{Fedoryuk-book}, which represents the base of the WKB
method. Notice that in matrix $M$, the energy $\mathcal{E}$ and  total angular momentum $J$ are the conserved quantities.

Following Ref.~\cite{Fedoryuk-book}, one writes
\begin{equation}
\xi(r)=\varphi(r) \exp\left[\frac{i}{\hbar}S(r)\right],
\end{equation}
which gives
\begin{equation}
\label{A-S-system}
\varphi'(r)=\frac{1}{\hbar}\left(M-i S'\cdot \sigma_0\right)\varphi(r),\quad S(r)=\int^r p(r)dr.
\end{equation}
Then one seeks the solution of system Eq.~(\ref{WKB-Dirac})
as an asymptotic series in powers of $\hbar$ (see, e.g., Ref.~\cite{Lazur2005TMP}):
\begin{equation}
p(r)=\sum\limits_{n=0}^\infty \hbar^n p_n(r),\quad \varphi(r)=\sum\limits_{n=0}^\infty \hbar^n\varphi^{(n)}(r).
\end{equation}
Substituting these expansions in Eqs.~(\ref{A-S-system}) and equating the coefficients of equal powers of $\hbar$ to zero, we
obtain an infinite system of recursive equations for the unknown scalar
$p_n(r)$ and vector $\phi^{(n)}(r)$ functions,
\begin{subequations}
\label{set-WKB}
\begin{align}
\label{zero-order}
& \! \! \! \left(M-i p_0I\right)\varphi^{(0)}(r)=0,\\
\label{high-orders}
& \! \! \! \left(M-i p_0I\right)\varphi^{(n+1)}(r)={\varphi^{(n)}}'(r)+i\sum\limits_{l=0}^np_{n+1-l}\varphi^{(l)}(r),
\end{align}
\end{subequations}
where $l=0,1,\ldots n$.
It follows from Eq.~(\ref{zero-order})  that $i p_0$ and $\varphi^{(0)}(r)$
must be the eigenvalues and eigenvectors of the matrix $M$.
In particular, we find that $p_0(r)=\pm p(r)$ with
\begin{equation}
\label{p(r)-electricfield}
p(r)=\frac{\hbar}{l}
\sqrt{\epsilon^2-\delta^2-j-\left(\frac{1}{4}-\beta^2\right)\rho^2-\frac{j^2}{\rho^2}-2\beta\epsilon\rho}.
\end{equation}
Recall that $\rho = r/l$.

The WKB approximation for the 2D massless Dirac fermions was discussed
in detail in Ref.~\cite{Zhang2012PRB}, so we directly proceed to the analysis of our problem. In particular,
the  Bohr-Sommerfeld  quantization condition for eigenenergies
for the Dirac fermions was obtained in  Ref.~\cite{Zhang2012PRB} both by using
Zwaan's method and by considering the condition of single
valuedness of the WKB wave function (see also
Refs.~\cite{Kormanyos2008PRB,Brack.book}). It reads
\begin{equation}
\label{BS-condition}
I(\epsilon,j)  \equiv  \int_{b}^{a}dr p(r)
= \pi\hbar \left(n_{\mathrm{BS}}+\frac{1-\theta(-j)}{2}\right),
\end{equation}
where $a \geq b$ are positive turning points [roots of the equation $p^2(r)=0$],
$n_{\mathrm{BS}}=0,1,2,\ldots$ is the Bohr-Sommerfeld quantum number,
$\theta(-j)$ is the step function  \cite{foot2}
whose presence takes into account the spinor nature of the
Dirac quasiparticles and the existence
of the lowest $n=0$ Landau level when the classically allowed region shrinks to a point.

\subsection{WKB approximation in the absence of electric field}
\label{sec:E=0}

Let us recapitulate how the WKB method can be applied in the absence
of an electric field, $E =0$. In this case, the integral $I(\epsilon,j)$ on the
left hand side (LHS) of the Bohr-Sommerfeld quantization condition Eq.~(\ref{BS-condition})
can be calculated:
\begin{equation}
\label{int-E=0}
I(\epsilon,j) =  \frac{\hbar \pi}{2}
(\epsilon^2 - \delta^2 - |j| - j).
\end{equation}
Solving Eq.~(\ref{BS-condition}) for the energy $\epsilon$, one recovers the
well-known Landau-level spectrum,
\begin{equation}
\label{LLspectrum-magfield}
\mathcal{ E}=\pm \sqrt{\Delta^2 + (2n_{\mathrm{BS}} + j+|j|+1-\theta(-j)) e H \hbar v_F^2/c}
\end{equation}
with $n_{\mathrm{BS}}=0,1,2,\ldots$.
Except for the asymmetric lowest Landau level,
this result  agrees perfectly with the exact solution
of the Dirac equation in the symmetric gauge (see Appendix~D of
Ref.~\cite{Gusynin2006PRB}, where the final result is written in
the form identical to the spectrum obtained
in the Landau gauge). For $j<0$, the spectrum Eq.~(\ref{LLspectrum-magfield})
is consistent with  the result in the Landau gauge if one identifies
the quantum number $n_{\mathrm{BS}}$ and  the Landau level index $n = n_{\mathrm{BS}}$.
To reproduce the spectrum for the $j>0$ case, one should relabel 
the quantum numbers
$2n =2 n_{\mathrm{BS}} + 2 j +1$, so  $n$ corresponds to 
the Landau-level index
and $\mathcal{ E}=\pm \sqrt{\Delta^2 + 2n  e H \hbar v_F^2/c}$
with $n=1,2, \ldots $ and $1/2 \leq j\leq n -1/2$.
Thus, one observes that in the absence of an electric field,
the WKB method reproduces well-known exact results (see, for example, Ref.~\cite{Zhang2012PRB}).

\subsection{WKB approximation in the radial electric field}
\label{sec:general}

Here we restrict ourselves to the gapless case,
$\Delta=0$. Then the expression Eq.~(\ref{p(r)-electricfield})
acquires the form
\begin{equation}
\label{p(r)-electricfield-roots}
p(r) =\frac{\hbar\sqrt{1/4-\beta^2}}{l\rho}\sqrt{-(\rho-\rho_1)(\rho-\rho_2)(\rho-\rho_3)(\rho-\rho_4)},
\end{equation}
where $\rho_i$ are the roots of the quartic equation $p^2(r) =0$:
\begin{equation}
\begin{split}
\rho_{1,2}&=\frac{-\epsilon\mp\sqrt{\epsilon^2-2j(1-2\beta)}}{1-2\beta},\\ \rho_{3,4}&=\frac{\epsilon\mp\sqrt{\epsilon^2-2j(1+2\beta)}}{1+2\beta}.
\end{split}
\end{equation}
As one can see, there are always two positive and two negative roots.
In particular, for the zero electric case considered in Secs.~\ref{sec:E=0} and \ref{sec:E=Delta=0}, the roots $\rho_1 = - \rho_4$ and $\rho_2 = - \rho_3$.  This
symmetry allows one to calculate the integral Eq.~(\ref{int-E=0}) using
the contour integration in the complex plane (see the corresponding integral
in the textbook \cite{Goldstein.book}).

Depending on the signs of $\epsilon$ and $j$,
these roots are ordered as follows: for $\epsilon>0$,
\begin{equation}
\label{roots>0}
\begin{split}
&\rho_1<\rho_2 <0< \rho_3< \rho_4 \quad \mbox{for} \quad j>0,\\
&\rho_1<\rho_3 < 0 < \rho_2<\rho_4 \quad \mbox{for} \quad j<0,
\end{split}
\end{equation}
and, for $\epsilon < 0$,
\begin{equation}
\label{roots<0}
\begin{split}
&\rho_3 < \rho_4 < 0 < \rho_1<\rho_2 \quad \mbox{for} \quad j>0,\\
&\rho_3 <\rho_2 < 0 <\rho_4 < \rho_2 \quad \mbox{for} \quad j<0.
\end{split}
\end{equation}
For $\beta \to \mp 1/2$ and $\epsilon \gtrless 0$,
the turning point $\rho_{4}$ ($\rho_{2}$)  moves to infinity
and some of the closed classical orbits transform into open trajectories.

We relabel these roots $a$, $b$, $c$, and $d$ and assume that they obey
the inequality $a>\rho>b>c>d$. Then the corresponding integral on the LHS
of Eq.~(\ref{BS-condition}) can be expressed in terms of complete Legendre
elliptic integrals of the first, $K(k)$, second, $E(k)$, and third, $\Pi(\nu,k)$, kinds.
In Appendix~\ref{sec:Appendix-int}, we obtain the following
result:
\begin{equation}
\label{int-general}
I(\epsilon,j)=-\frac{\hbar}{ \sqrt{1-4\beta^2}}\left[\beta\epsilon J_2-(\epsilon^2-j)J_1+2j^2J_{-1}\right],
\end{equation}
where
\begin{subequations}
\label{Js}
\begin{align}
\label{J1}
J_{1} =& \frac{2}{\sqrt{(a-c)(b-d)}}\left[(b-c)\Pi\left(\frac{a-b}{a-c},k\right)
+c K(k)\right], \\
\label{J-1}
J_{-1}= &\frac{2}{\sqrt{(a-c)(b-d)}}\left[\frac{c-b}{b c}\Pi\left(\frac{c(a-b)}{b(a-c)},k\right)+\frac{1}{c} K(k)\right],\\
\label{J2}
J_2=&\frac{1}{\sqrt{(a-c)(b-d)}}\bigg[ \left(-a(b-c)+c(b+c)\right)K(k)\nonumber\\
&+\left.(a-c)(b-d)E(k) \right.\\
&+ \left. (b-c)(a+b+c+d)\Pi\left(\frac{a-b}{a-c},k\right)\right],\nonumber
\end{align}
\end{subequations}
with
\begin{equation}
\label{k}
k^2=\frac{(a-b)(c-d)}{(a-c)(b-d)}.
\end{equation}
The quantization condition Eq.~(\ref{BS-condition}) with the LHS given by  Eq.~(\ref{int-general}) represents a transcendental equation for energies
of bound states in terms of complete elliptic integrals.
This complicated equation cannot be solved explicitly for the energy.
However, they can be calculated efficiently with numerical methods.
It is shown in Appendix~\ref{sec:E=Delta=0} that the spectrum in the zero
electric field, $E=0$, limit can be directly derived from
Eq.~(\ref{int-general}).

\section{Results}
\label{sec:results}

\subsection{The gapless  case, $\Delta=0$}
\label{sec:gappless}

\begin{figure}[!h]
\includegraphics[width=.45\textwidth]{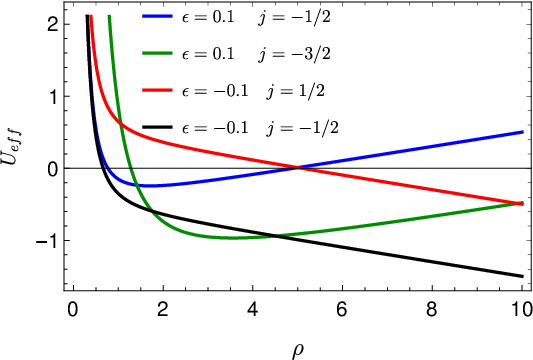}
\caption{The effective potential $U_{\mathrm{eff}}(\rho)$
versus $\rho$ for $\beta =1/2$ and
different values of $\epsilon$ and $j$.
}
\label{fig-potential}
\end{figure}

To develop a qualitative understanding, it is useful to represent the
momentum Eq.~(\ref{p(r)-electricfield}),
\begin{equation}
p(r) = \frac{\hbar}{l} \sqrt{\epsilon_{\mathrm{eff}} - U_{\mathrm{eff}} },
\end{equation}
in terms of the effective WKB energy $\epsilon_{\mathrm{eff}} \equiv \epsilon^2$
(we set $\delta =0$)
and the effective potential:
\begin{equation}
\label{effective-potential}
U_{\mathrm{eff}}(\rho) \equiv \left(\frac{1}{4} - \beta^2 \right)\rho^2 + 2 \beta \epsilon \rho  + \frac{j^2}{\rho^2} + j.
\end{equation}
It is clear from Eqs.~(\ref{p(r)-electricfield-roots})--(\ref{roots<0}) that the classically
allowed region is situated between the positive roots denoted above as $a>\rho > b >0$.
One can see  that a finite motion and quantized energy levels are possible for $|\beta| < 1/2$ when the potential Eq.~(\ref{effective-potential}) grows 
as $\rho \to \infty$.
Furthermore, for $\beta = \pm 1/2$ the character
of motion becomes dependent on the linear in $\rho$ and constant terms
of   the potential Eq.~(\ref{effective-potential}) (see Fig.~\ref{fig-potential}).
We return to this point below.

For comparison we recapitulate that for the problem with the repulsive Coulomb
potential $V(r) = \hbar v_F \mathfrak{g} /r$ with dimensionless coupling $\mathfrak{g}$
\cite{Zhang2012PRB} (see also Refs.~\cite{Khalilov2000PRA,Gamayun2011PRB,Li2021})
the effective potential
\begin{equation}
\label{effective-potential-Coul}
U_{\mathrm{eff}}^{C}(\rho) \equiv  \frac{\rho^2}{4} +
\frac{2\mathfrak{g} \epsilon}{\rho}  + \frac{j^2 -\mathfrak{g}^2}{\rho^2} + j.
\end{equation}
In contrast to the case considered in this work, the effective potential
for the Coulomb interaction is always positive as $\rho \to \infty$ and the behavior
of the system is determined by its dependence for $\rho \to 0$. The critical value
of $\mathfrak{g}$ when $U_{\mathrm{eff}}^{C}(\rho) $ becomes negative turns out
to be dependent on the absolute value of total angular momentum quantum number $j$.

\subsubsection{Numerical solution in WKB approximation}
\label{sec:WKBresults}

\begin{figure*}[!htb]
\includegraphics[width=.99\textwidth]{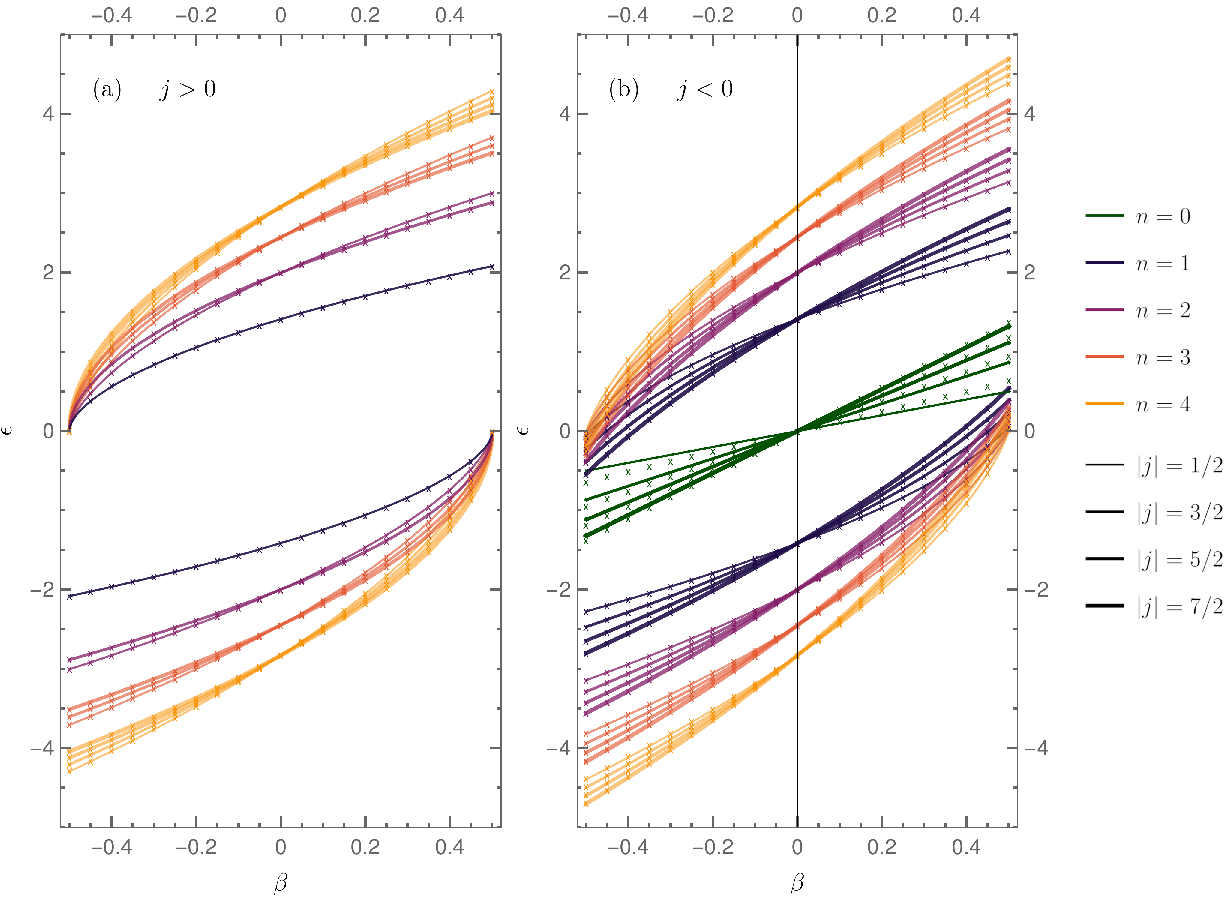}
\caption{WKB  and exact diagonalization (ED) spectra in units
$\epsilon_0 = \hbar v_F/l$ versus electric field in terms of $\beta = c E/(v_F H)$ for the gapless, $\Delta=0$ case.
Panel~(a) shows positive values of $j$ and panel~(b) shows negative values of j.
In both panels, solid lines show the results of the WKB approximation and crosses
show the results calculated using ED.
The Landau levels with $n=0,1,2,3,4$ are shown by the green,
dark purple, purple lines, red and orange lines, respectively.
The levels with $|j| =1/2, 3/2,5/2,7/2$ are marked by the increasing
thickness of the lines.
}
\label{fig-spectra}
\end{figure*}
The Bohr-Sommerfeld quantization condition Eq.~(\ref{BS-condition})
with the LHS given by Eq.~(\ref{int-general}) represents the
transcendental  WKB equation for the energy spectrum. Its
numerical solution describing the dependence of the energy $\mathcal{E} $
in units $\epsilon_0 = \hbar v_F/l$ versus electric field in terms of
the dimensionless parameter $\beta$ is shown in Fig.~\ref{fig-spectra}.
For better readability, we plotted the cases $j>0$ and $j <0$ on
the separate left~(a) and right~(b) panels, respectively.

For $j >0$, we have done the relabeling of the Landau levels as described
below Eq.~(\ref{LLspectrum-magfield}),
so  the notation $n=1,2,3,4$ corresponds to the Landau-level
index.  The lowest $n=0$ level is present only for $j<0$.

First, we observe that the presence of a finite electric field removes
the degeneracy of the Landau levels with the different
total angular momenta $|j| =1/2,3/2,5/2, 7/2$ that are marked
by the increasing thickness of the lines, respectively.
As discussed  below Eq.~(\ref{LLspectrum-magfield}), for $j>0$
to keep  $n$ being the Landau-level index,
one should set a restriction on the
allowed values of  the total angular momentum
quantum number $1/2 \leq j\leq n -1/2$. Thus, for $n=1$ there is only one
line with $j=1/2$; for $n=2$ there are two lines with $j=1/2,3/2$ and so on.
On the other hand, there is no such restriction for the negative values of $j$
and we just restricted ourselves by showing the lines with $j=-1/2,-3/2, -5/2,-7/2$.
The apparent asymmetry in the allowed range of the values of $j$ reflects that  circulating
with positive $j$ in the presence of a
magnetic field directed in the positive $z$ direction costs energy for an electron,
whereas circulating with negative $j$ does not \cite{Bhuiyan2020AJP}.

For $E=0$, these levels become degenerate and their
energies are in agreement with Eq.~(\ref{LLspectrum-magfield}).
One can see that the dependencies  $\epsilon(\beta)$ for the electron-like
levels are symmetric with respect to the coordinate origin,
$\epsilon(\beta) = - \epsilon(-\beta)$, as compared
to the corresponding dependencies  $\epsilon(\beta)$ for the hole-like
levels in both panels [see Eq.~(\ref{effective-potential-Coul})].

An interesting property of the presented results is that the
character of the dependence of the spectra on $\beta$ turns out to
depend on the sign of $j$. The hole-like Landau levels
with different values of $j >0$ and given $n$ are non-degenerate for $\beta = -1/2$.
Then, as $\beta$ increases the distance between these levels diminishes and
they become degenerate as $\beta$ reaches zero. Then, as $\beta$ becomes positive,
the distance between the levels starts to increase, but when $\beta$ grows further
the distance between the levels diminishes and all levels with different $j>0$ and
$n \geq 1$ collapse to zero energy for $\beta =1/2$.
The behavior of the electron-like levels is consistent with above-mentioned
symmetry, so  their collapse occurs as $\beta$ decreases from $1/2$
to $-1/2$.

This critical value $\beta =1/2$ first appears in Ref.~\cite{Sun2014PRB}, where Coulomb
impurity spectra under external electric and magnetic fields were
studied using the coupled series expansion method.
Although the main focus of Ref.~\cite{Sun2014PRB} is on the Coulomb impurity,
it also contains some numerical results corresponding to
the field configuration considered here.

The behavior of levels with $j < 0$ is drastically different.
First, these levels include the $n=0$ Landau level
whose energy depends linearly on the electric field.
The hole-like levels for $-1/2 \leq \beta \leq 0$ behave similarly
to the positive $j$ case. However,  for $0 \leq \beta \leq 1/2$
the level distances increase and as $\beta$ gets closer to $1/2$
the levels with $j = -7/2,-5/2,-3/2,-1/2$ one by one cross zero and
have positive constant values at $\beta=1/2$.
There is no level collapse in this case.
The behavior of the electron-like levels is consistent with the above-mentioned
symmetry, so the level distances increase and their energies
become negative as $\beta$ decreases from $1/2$ to $-1/2$.

The values of $\beta$ for which the negative-$j$ levels intersect the zero
energy line can be found analytically. Indeed, the function $I(\epsilon =0,j)$
defined by Eq.~(\ref{BS-condition}) acquires a rather simple form
\begin{equation}
\label{I-zero-energy}
I(\epsilon =0,j)=
\hbar\int_{\rho_1}^{\rho_2} d \rho
\sqrt{-(1/4-\beta^2)\rho^2-\frac{j^2}{\rho^2}-j},
\end{equation}
with $\rho_{\pm} = \sqrt{ (-2j)/(1 \pm 2 |\beta|)}$.
Evaluating the integral Eq.~(\ref{I-zero-energy}) and substituting
the result in Eq.~(\ref{BS-condition}), we arrive at the equation
that determines the intersection points $\beta=\beta(\epsilon=0,n,j)$,
which can be solved explicitly:
\begin{equation}
\label{beta-e=0}
\beta(n,j)= \pm\frac{\sqrt{n(n+|j|)}}{2n+|j|}, \quad j<0,\,\, n=0,1,\ldots\, .
\end{equation}
Here we took into account that $n=n_{\mathrm{BS}}$ for $j<0$. Since
the nontrivial solutions Eq.~(\ref{beta-e=0}) exist for all allowed values of $n$
and $j <0$, this proves that all corresponding levels cross zero energy.

As mentioned above, the levels $\epsilon(n,j,\beta)$
approach the bound-state energies $\epsilon(n,j,\beta=1/2)$
as $\beta$ reaches the value $1/2$.
We have also obtained Eq.~(\ref{BSQ-beta=1/2}) (see Appendix~\ref{sec:Appendix-int-beta=1/2}) for these energies $\epsilon(n,j,\beta=1/2)$ and found its approximate analytic solution:
\begin{equation}
\label{energies-beta=1/2}
\epsilon (n,j,\beta=1/2) \approx \left(\frac{|j|}{2}-\sqrt{\frac{3\sqrt{3}n|j|}{2}}\right)^{1/2},\quad |j|\gg \pi n.
\end{equation}

\begin{figure}[!h]
\includegraphics[width=.45\textwidth]{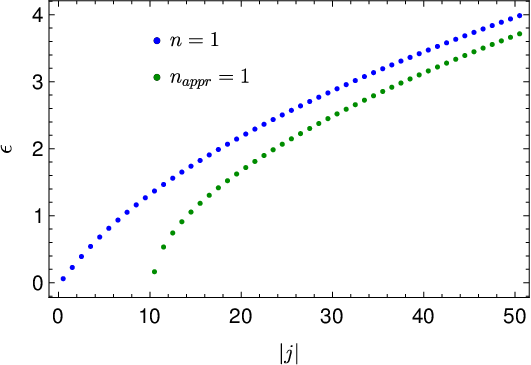}
\caption{The WKB eigenenergies
$\epsilon (n,j, \beta =1/2)$ versus the total angular momentum
quantum number $j$ for $n=1$. Large blue dots correspond to
the numerical solution of Eq.~(\ref{BSQ-beta=1/2}) and small green dots
are for the approximate solution Eq.~(\ref{energies-beta=1/2}).
}
\label{fig:j-plot}
\end{figure}

The numerical solution of Eq.~(\ref{BSQ-beta=1/2})
for the eigenenergies
$\epsilon (n,j, \beta =1/2)$ versus the total angular momentum
quantum number $j$ is shown in Fig.~\ref{fig:j-plot} by the large blue dots.
The approximate solution Eq.~(\ref{energies-beta=1/2}) is plotted for comparison
(small green dots).
One can see that the approximate solution has a good agreement
with the numerical one for large values of $j$.

As follows from Figs.~\ref{fig-spectra}, \ref{fig:j-plot}, and analytic consideration,
for $\beta =1/2$ there are
bound states for positive energies, $\epsilon$, irrespective to
the sign of $j$, and bound states are absent for  negative energies,
$\epsilon$. To explain this,
we plotted in Fig.~\ref{fig-potential} the effective potential
Eq.~(\ref{effective-potential})
for $\beta = 1/2$ and different values of $\epsilon$ and $j$.

As mentioned above for $\beta =1/2 $,
the behavior of the potential is governed by
the last three terms of Eq.~(\ref{effective-potential}).
For $\epsilon >0$, the effective potential
$U_{\mathrm{eff}}(\rho)$ grows linearly as $\rho \to \infty$, so 
the quasiparticle orbits remain closed and correspond to the bound states.
For $\epsilon < 0$, the effective potential
$U_{\mathrm{eff}}(\rho)$  decreases linearly as $\rho \to \infty$,
so  there are no closed orbits and bound states.
Since for $j <0$ the solutions for $\beta =1/2$ have the positive energy
and thus are the bound states, there is no Landau-level collapse
in this case.

On the other hand, for $j>0$ all levels with negative energy
merge to one point and collapse (see Fig.~\ref{fig-spectra} left panel).
It is important to stress that because for $j>0$
the expression under the square root in the integrand of Eq.~(\ref{I-zero-energy})
is negative, Eq.~(\ref{BS-condition}) does not have a solution in this case.
Thus, in contrast to the case of the crossed uniform magnetic and electric fields,
in the studied geometry the collapse points with $\epsilon =0$ and $\beta = \pm 1/2$
do not belong to the spectra.

\subsubsection{Diagonalization and shooting method}
\label{sec:EDDresults}

To verify the accuracy of the WKB approximation, we use two numerical methods: discretization with exact diagonalization and shooting method.

For the diagonalization method, we use the Hamiltonian which one can obtain from Eq.~(\ref{matrix-M}). To do this, one should rewrite Eq.~(\ref{matrix-M})
in terms of an eigenvalue problem multiplying it by $i\sigma_2$ matrix.
The first derivative (momentum) is discretized using finite differences.
In this way, the eigenvalue problem can be solved on a 1D lattice with
two orbitals, where the momentum is turned into hopping terms and
the other parts of the Hamiltonian into on-site terms.

It is worth noting that the choice of grid spacing of the lattice
should correspond to the behavior of the onsite term to reach all
important values. To increase the efficiency, we use a non-constant grid
with slowly increasing spacing, thus obtaining more dots closer to zero, where our effective potential is singular.

To implement the discretization method, the {\large K}WANT \cite{kwant} Python package was used. With this package, we build the Hamiltonian of a 2000-site-long chain and then numerically diagonalize it.

It also should be noted that with such a method one can obtain
a fermionic-doubling  effect  \cite{NielsenNin}, as we
observe in our numerical work.
Although this effect is not clearly seen in the gapless case,
it immediately appears if one introduces a non-zero gap,
thus resulting in the spectra doubling.

The shooting method is an integration process performed on the system Eq.~(\ref{matrix-M}) with a guessed energy-parameter (for more details, see
Appendix~\ref{sec:shooting}). Such a method does not
result in the fermionic doubling.
Thus we can numerically solve  the corresponding equations for each
$\mathbf{K}_{\pm}$ separately and prove that the diagonalization method
produces the results for the two valleys together.

The exact diagonalization method is more efficient in a sense of speed
and precision in comparison with shooting. Thus, we use the diagonalization as a main computation method and the shooting method as an auxiliary one to distinguish different $\mathbf{K}_{\pm}$ valleys.

The results of the numerical methods described above for the gappless case
are shown in Fig.~\ref{fig-spectra} by crosses. The comparison with WKB (solid
lines in Fig.~\ref{fig-spectra}) shows that the WKB approximation is in a good
quantitative agreement with the numerical calculations for the whole range of the values of $\beta$, $j$, and $n$. There are only small quantitative discrepancies in the energies of the lowest, $n = 0$, Landau level seen in Fig.~\ref{fig-spectra}(b). However, this deviation decreases with increasing $|j|$.
The gapped case was investigated using numerical methods
only and is considered in what follows.

\subsection{The gapped case}
\label{sec:gapped}

\begin{figure*}[!htb]
\includegraphics[width=.95\textwidth]{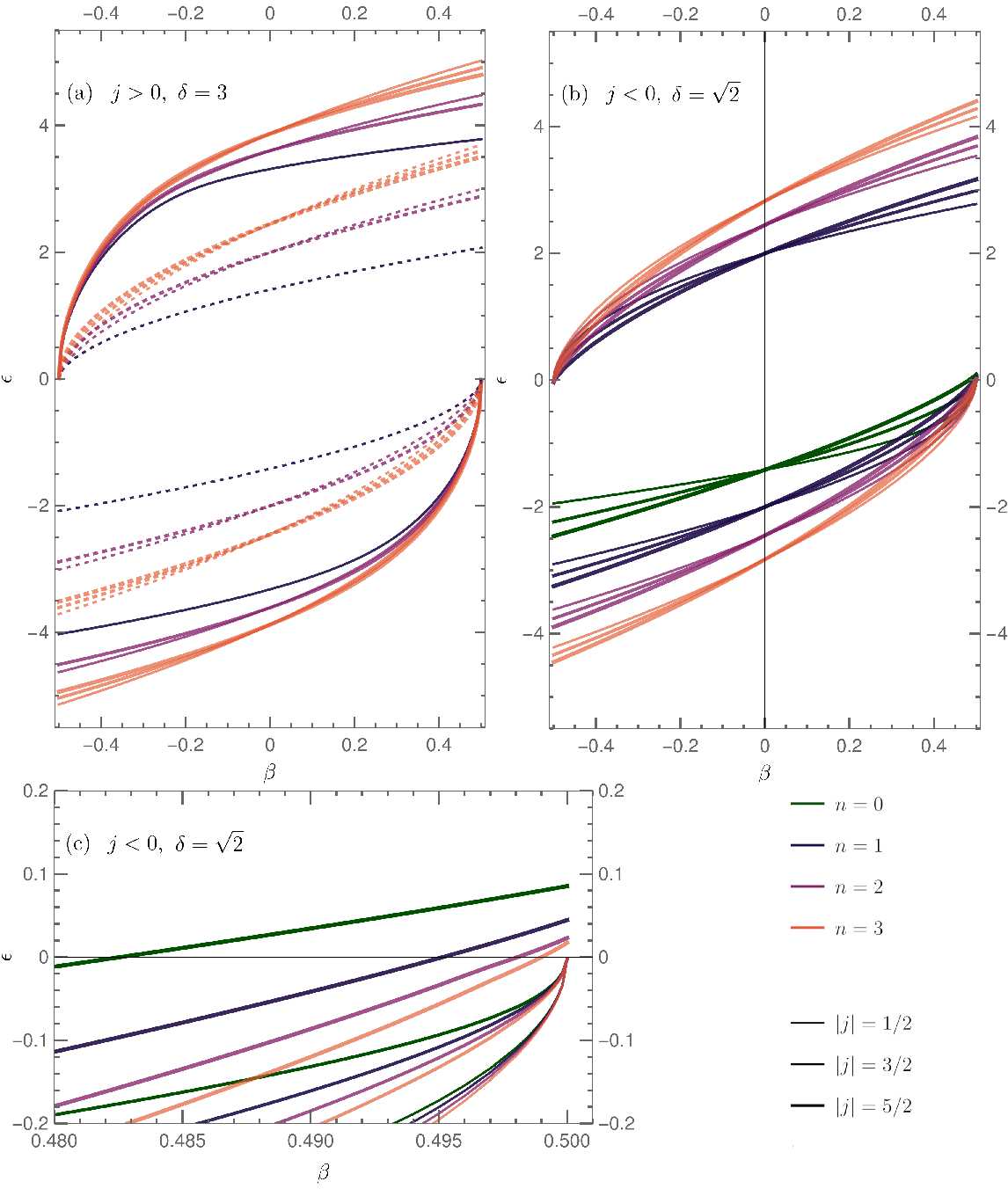}
\caption{Spectra for the $\mathbf{K}_{+}$ point obtained
by the shooting method in units $\epsilon_0 = \hbar v_F/l$ versus
electric field in terms of $\beta = c E/(v_F H)$ for the finite gap,
$\delta =  l \Delta/(\hbar v_F) $, case.
(a) Positive $j$: The solid lines are the solutions for
$\delta =3$ and  the dashed lines
are for $\delta =0$.
(b) Negative $j$: The gap $\delta = \sqrt{2}$.
(c) The zoom of  plot  (b) in the vicinity of $\beta=1/2$.
The same color and thickness scheme as the one displayed in Fig.~\ref{fig-spectra}
is used. }
\label{fig:5}
\end{figure*}
Now we turn to the finite $\Delta$ case. A crucial feature is that
the fully quantum mechanical description of the problem,
as one can see from the second-order Eq.~(\ref{2nd-order})
[see also Eqs.~(\ref{Dirac-system-WKB}) and (\ref{N})]
is sensitive to the sign of $\Delta$ when $\beta$ is nonzero.
As mentioned above Eq.~(\ref{derivative-angle}), the solution at
the $\mathbf{K}_-$ point is obtained from the solution at
the $\mathbf{K}_+$ point by changing $\Delta \to - \Delta$.
Thus, for  finite values of gap $\Delta$ and electric field $E$,
the results for the spectra are expected to be valley dependent.

On the other hand, one can see from Eq.~(\ref{p(r)-electricfield})
that even for a finite $\delta$ the WKB approximation does not
distinguish the valley, because it contains only $\delta^2$.

The fully numerical solution obtained using the methods described in
Sec.~\ref{sec:EDDresults} for the dependence of the energy $\mathcal{E} $
in units $\epsilon_0 = \hbar v_F/l$ versus electric field in terms of
the dimensionless parameter $\beta$ in the finite $\delta$ case
is shown in Fig.~\ref{fig:5}.
We used the same color and thickness scheme as the one displayed
in Fig.~\ref{fig-spectra}, but for clarity of the figure we do
not include the $n=4$ level and $|j| =7/2$ (see also  Supplemental Material \cite{suppl}).

Similarly to Fig.~\ref{fig-spectra},
cases $j > 0$ and $j < 0$ are shown on the separate
(a) and (b) panels, respectively. However, in~Fig.\ref{fig:5}~(a),
we selected a larger value of $\delta =3$ to make more
distinct the difference between the gapped (solid lines)
and gapless (dashed lines) cases. In both cases
the levels with different $j > 0$ and $n\geq 1$ collapse to
zero energy for $|\beta| = 1/2$.

One can see that in the contrast to the gapless case (see Fig.~\ref{fig-spectra}),
the dependencies  $\epsilon(\beta)$ for the energy levels
levels are no more symmetric with respect to the coordinate origin.

Yet, in the absence of an electric field the positive and negative energy levels
are symmetric with respect to the zero energy except for the lowest $n=0$
Landau level \cite{Gusynin2006PRB}.
Since as discussed below Eq.~(\ref{LLspectrum-magfield}) for $j>0$,  
the Landau-level index takes the values $n=1,2, \ldots,$ the lowest level 
is absent in Fig.~\ref{fig:5}~(a), so for $\beta=0$
all electron- and hole-like levels are symmetric with respect to
the origin. The lowest Landau level is
present in Figs.~\ref{fig-spectra}~(b) and \ref{fig:5}~(b).
Notice that for the chosen $\mathbf{K}_+$ point and direction of
the magnetic field, the energy of this level has to be $-\Delta$ for $E=0$
\cite{Gusynin2006PRB}. This agrees with Fig.~\ref{fig:5}~(b),
where the Landau levels for only the $\mathbf{K}_+$ point are shown.
[Recall that Fig.~\ref{fig:5}~(a) also contains the Landau levels for $\mathbf{K}_+$
point and there is no difference between the $\mathbf{K}_\pm$ points
in the gapless case considered in Sec.~\ref{sec:gappless}.]

This asymmetry of the energy of the lowest Landau level for $\beta =0$
would disappear if one takes into consideration the second
$\mathbf{K}_-$ point with the lowest level having a positive energy $\Delta$
for $E=0$. Note that the general asymmetry between  the electron- and hole-like
levels with respect to the coordinate origin for $\delta, \beta \neq 0$
would also disappear when the levels for $\mathbf{K}_-$ point
are considered. It is worth mentioning that as follows from
Eq.~(\ref{LL-collapse}) this valley asymmetry is absent in
the crossed uniform magnetic and electric fields configuration.

To obtain Fig.~\ref{fig:5}~(b), we took a smaller value of the gap
$\delta =\sqrt{2}$ such that $|j| < \delta^2 $ for $j = -1/2,-3/2$
and $|j| > \delta^2 $ for $j =-5/2$. To resolve the $\epsilon(\beta)$
dependencies in the vicinity of $\beta =1/2$, a zoom of
this region is shown in Fig.~\ref{fig:5}~(c).
One can see that the levels with $j=-1/2,-3/2$
collapse to  zero energy, while the levels with $j=-5/2$
cross zero and approach positive constant values. The evolution
of Landau levels with the increase of the gap is presented in
the Supplemental Material \cite{suppl}.

Generalizing the analysis of Eq.~(\ref{I-zero-energy}) for the finite
$\delta$ case, one finds that the energy levels may cross the
$\epsilon=0$ line if $|j| > \delta^2$. The levels with $|j| < \delta^2$
cannot cross this line and thus approach the $\epsilon=0$ point
for $|\beta| =1/2$.

This behavior may also be understood qualitatively by
considering Fig.~\ref{fig-potential} for the effective potential plotted
for various values of $j$. The classically allowed region is determined
by the points where $U_{\mathrm{eff}}(\rho)$ crosses the
effective energy  $\epsilon_{\mathrm{eff}} = \epsilon^2- \delta^2$.
For $\delta =0$ and $\beta =1/2$,
the positive-$j$ levels collapse
and the negative-$j$  levels correspond to the bound
states with $\epsilon >0$. However, the presence of a finite $\delta$
shifts down the values of $j<0$ allowed for the collapse  by
$-\delta^2$.

Thus, the presence of the gap extends
the values of $j$ allowed for the Landau level
collapse from positive-$j$ to
the negative-$j$ levels with $|j| < \delta^2$.

\section{Conclusion}
\label{sec:conclusion}

As  already mentioned in the Introduction, the Landau-level collapse was
already observed experimentally \cite{Singh2009PRB,Gu2011PRL}.
For example, in Ref.~\cite{Gu2011PRL}
the Shubnikov-de Haas-type resonances arising
from quantized states associated with closed orbits are
used to directly observe the competition between magnetic
confinement and deconfinement due to electric field.
While these observations were made in the rectangular geometry,
there are no limitations for repeating them in the
circular Corbino geometry considered in this work.

High-mobility Corbino devices in a dual-gated geometry were recently studied
in Ref.~\cite{Dean2019PRL}.  Bulk conductance measurement outperforms previously reported Hall bar measurements  and allowed one to observe both the integer and fractional QHE states.
It should be technically possible to modify the existing devices
introducing the radial electric field by gating. Another way to apply
the radial electric field might be achieved by injecting a high current
density in the device \cite{Singh2009PRB}.
Since for nonzero electric field the degeneracy of
the Landau levels with different total angular momenta
$j$ is lifted, this should be manifested in the
transport measurements. Another possibility to observe
the predicted features is to employ scanning tunneling  spectroscopy
that allowed one to observe Dirac Landau levels in graphene \cite{Li2007NP}.

Further increase of the
electric field would allow one to realize
the Landau-level collapse in the geometry studied in
this paper. Moreover,
the critical electric field required for the collapse
is $\pm E_c/2 $  is twice smaller than
the corresponding field $E_c= (v_F/c)H$
for the rectangular geometry that should make its observation easier.
In contrast to the rectangular geometry, the
collapse in the circular geometry is sensitive to the sign of the electric
filed, so  the hole (electron)-like levels collapse at $ +(-) E_c/2$.
One can estimate that for the magnetic field
$H =\SI{1}{T}$ and the Fermi velocity $v_F = \SI{1e6}{m/s}$,
the critical electric field in the circular geometry
is $E_c = \SI{0.5e4}{V/cm}$ which should be possible to create
by gating.

It should also be possible to investigate the collapse
in the gapped case by using graphene
placed on top of hexagonal boron nitride (G/hBN).

Another experimental setup that would allow one to realize the Landau-level collapse
by generating strain induced either pseudomagnetic or electric fields was
suggested in Refs.~\cite{Castro2017PRB,Grassano2020PRB}, respectively.
Although the corresponding experiments were not done, it should
be possible to make them both in rectangular and Corbino geometries.
Finally, it might also be necessary to generalize the
presented results with a radial electric field for a
finite-size  Corbino disk \cite{Yerin2021PRB}.


\begin{acknowledgments}
We would like to thank the Armed Forces of Ukraine for providing security to perform this work.
I.O.N. and V.K. would like to thank I.C.~Fulga and J.~van~den~Brink for fruitful discussions.
I.O.N., V.P.G. and S.G.Sh. are grateful to O.O.~Sobol for valuable  discussion.
V.P.G. and S.G.Sh. thank A.A.~Varlamov for useful
remarks. I.O.N., V.P.G. and S.G.Sh. acknowledge the support by  National Research Foundation of Ukraine (NRFU) Grant  No.2020.02/0051, "Topological phases of matter and excitations in Dirac materials, Josephson junctions and magnets" in the period 2020-2021. The continuation of the funding in 2022 by
NRFU became impossible due to the Russian war against Ukraine.

\end{acknowledgments}	


\appendix

\section{Calculation of the integral $I(\epsilon,j)$ for  finite $\beta$}
\label{sec:Appendix-int}

We calculate the integral (\ref{int-general})
\begin{equation}
I(\epsilon, j) = \sqrt{1/4-\beta^2} S,
\end{equation}
with
\begin{equation}
\label{S}
S =\int_b^a\frac{dx\, y(x)}{x},
\end{equation}
and $y(x)=\sqrt{(a-x)(x-b)(x-c)(x-d)}$. Here we introduced notations
of Ref.~\cite{Bateman3}, where similar integrals are calculated.
Multiplying the numerator and denominator of the integrand in Eq.~(\ref{S}) by $y(x)$,
one obtains
\begin{equation}
\label{S1}
S=\int\limits_b^a\frac{a_0x^3+4a_1x^2+6a_2x+4a_3+a_4/x}{y(x)},
\end{equation}
where $y^2(x)= a_0x^4+4a_1x^3+6a_2x^2+4a_3x+a_4$ with
\begin{equation}
\begin{split}
a_0 &=-1,\qquad 4a_1=a+b+c+d,\\
6a_2 &=-a(b+c)-b(c+d)-d(a+c),\\
4a_3 &= acd+bcd+abc+abd,\qquad a_4=-abcd.
\end{split}
\end{equation}
Accordingly, for the relabeled roots $\rho_4\to a,\rho_3\to b,
\rho_2\to c,\rho_1\to d$
(recall that we assumed that $a>x>b>c>d$), one obtains that
\begin{equation}
\label{ai}
\begin{split}
a_0 & =-1, \quad a_1  =-\frac{2\beta\epsilon}{1-4\beta^2}, \quad
a_2=\frac{2(\epsilon^2-j)}{3(1-4\beta^2)}, \\
a_3 & =0,  \qquad
a_4=-\frac{4j^2}{1-4\beta^2}.
\end{split}
\end{equation}
Defining the integrals
\begin{equation}
J_n=\int\limits_b^a\frac{dx\,x^n}{y(x)},\quad n=-1,0,1,2,3,
\end{equation}
one can rewrite Eq.~(\ref{S1}) in the following form:
\begin{equation}
\label{S2}
S=a_0J_3+4a_1J_2+6a_2J_1+4a_3J_0+a_4J_{-1}.
\end{equation}
To get rid of the integral $J_3$ from Eq.~(\ref{S2}), we consider
the derivative
\begin{equation}
\frac{dy}{dx}=\frac{2a_0x^3+6a_1x^2+6a_2x+2a_3}{y},
\end{equation}
and integrating it $b$ to $a$ we obtain the identity
\begin{equation}
a_0J_3+3a_1J_2+3a_2J_1+a_3J_0=0.
\end{equation}
Hence
\begin{equation}
S=a_1J_2+3a_2J_1+3a_3J_0+a_4J_{-1},
\end{equation}
and using the coefficients Eqs.~(\ref{ai}) we arrive at the expression
\begin{equation}
S=-\frac{2}{1-4\beta^2}\left[\beta\epsilon J_2-(\epsilon^2-j)J_1+2j^2J_{-1}\right].
\end{equation}
The integrals $J_1$ and $J_{-1}$ given by Eqs.~(\ref{J1}) and (\ref{J-1}), respectively,
can be found in Ref.~\cite{Gradshtein.book} [see Eqs.~(3.148.6) and (3.150.4)].
The integral $J_2$ is considered, for example, in Ref.~\cite{Lazur2008TMP},
where it is rewritten in the form
\begin{equation}
\label{J2-T2}
\begin{split}
J_2&= \frac{2}{\sqrt{(a-c)(b-d)}}\bigg[ c^2 K(k)
 +2c(b-c) \\
& \times \Pi\left(\frac{a-b}{a-c},k\right)
+(b-c)^2T_2\left(\frac{a-b}{a-c},k\right)\bigg],
\end{split}
\end{equation}
where
\begin{equation}
\label{T2}
T_2(\nu,k)=\int\limits_0^{\pi/2}\frac{d\varphi}{(1-\nu\sin^2\varphi)^2\sqrt{1-k^2\sin^2\varphi}}.
\end{equation}
Evaluating the last integral (see Eq.~(2.592.6) in Ref,~\cite{Gradshtein.book}),
one obtains
\begin{equation}
\label{T2-fin}
\begin{split}
T_2(\nu,k)& =-\frac{1}{2(1-\nu)}K(k)
 -\frac{\nu}{2(k^2-\nu)(1-\nu)}E(k) \\
&+\frac{k^2(3-2\nu)+\nu(\nu-2)}{2(k^2-\nu)(1-\nu)}\Pi(\nu,k).
\end{split}
\end{equation}
Finally, substituting (\ref{T2-fin}) in Eq.~(\ref{J2-T2}),
we obtain  Eq.~(\ref{J2}).

\section{Calculation of the integral $I(\epsilon,j,\beta=1/2)$ and energies $\epsilon(n,j,\beta=1/2)$}
\label{sec:Appendix-int-beta=1/2}

We provide below the results of the analytical consideration
of the $\beta =1/2$ case. Formally, an equation for the energies
$\epsilon(n,j,\beta=1/2)$ is still given by the
Bohr-Sommerfeld quantization condition
Eq.~(\ref{BS-condition}), but its LHS is simpler than
the generic expression in Eq.~(\ref{int-general}).
Indeed, the expression for the momentum Eq.~(\ref{p(r)-electricfield})
acquires a more simple form than Eq.~(\ref{p(r)-electricfield-roots}),
\begin{equation}
\label{p(r)-beta=1/2}
p(r)=
\frac{\hbar}{l\rho}\sqrt{-\epsilon(\rho-\rho_1)(\rho-\rho_2)(\rho-\rho_3)},
\end{equation}
where the roots  of the cubic equation $p^2(r) =0$ are
\begin{equation}
\label{roots-beta=1/2}
\rho_{1,2}=\frac{\epsilon\pm\sqrt{\epsilon^2-4j}}{2},\quad \rho_3=-\frac{j}{\epsilon}.
\end{equation}
Again depending on the signs of $\epsilon$ and $j$, these roots
are ordered differently.

The calculation of $I(\epsilon, j,\beta =1/2)$ is rather
similar to the general case considered in Appendix~\ref{sec:Appendix-int}
and we restrict ourselves by showing the final result.
We obtain
\begin{equation}
\label{int-beta=1/2}
I(\epsilon, j,\beta =1/2) =
\frac{\hbar}{\sqrt{\epsilon}}\left(\frac{\epsilon^2-j}{3}L_1-j^2 L_{-1}\right),
\end{equation}
where
\begin{equation}
L_n =
\int_b^a\frac{dx x^n}{\sqrt{(a-x)(x-b)(x-c)}},   \quad a>b>c,
\end{equation}
with $n =-1,1.$
Using  Eqs.~(3.132.5) and (3.137.6) from Ref.~\cite{Gradshtein.book}, we have
\begin{subequations}
\label{Ls}
\begin{align}
\label{L1}
L_{1}= &\frac{2c}{\sqrt{a-c}}K(k)+2\sqrt{a-c}E(k),\\
\label{L-1}
L_{-1}= & \frac{2}{a\sqrt{a-c}}\Pi\left(\frac{a-b}{a},k\right),
\end{align}
\end{subequations}
with $k^2=(a-b)/(a-c)$.

In the considered $\epsilon>0$ and $j<0$ case, we have
$a = \rho_3 = |j|/\epsilon$, $b= \rho_1$ and $c= \rho_2$.
Substituting Eq.~(\ref{int-beta=1/2}) in the quantization condition Eq.~(\ref{BS-condition})
we arrive at the following transcendental equation for the eigenenergy
$\epsilon(j,n,\beta =1/2)$:
\begin{equation}
\label{BSQ-beta=1/2}
\begin{split}
\frac{\sqrt{2}}{3 \sqrt{t s} } & \left[ (t+1) s E (k)
+ (t+1)(t-\sqrt{t^2+4t}) K (k) \right. \\
& \left. - 6 t \Pi(\nu, k) \right] = \frac{\pi n}{|j|},
\end{split}
\end{equation}
where we introduced $t=\epsilon^2/|j|$  and
\begin{equation}
\begin{split}
s  = 2-t+\sqrt{t^2+4t}, & \qquad
r =  2-t-\sqrt{t^2+4t},\\
\nu = 1-\frac{t+\sqrt{t^2+4t}}{2}, & \qquad k^2 = \frac{r}{s}.
\end{split}
\end{equation}
In the limit $|j|\to\infty$, Eq.~(\ref{BSQ-beta=1/2})
has an exact solution $t=1/2$, so asymptotically
$\epsilon\sim\sqrt{|j|/2}$. Expanding the LHS of Eq.~(\ref{BSQ-beta=1/2})
around the point $t=1/2$ up to $(t-1/2)^2$ term one obtains
the following equation:
\begin{equation}
\frac{2\pi}{3\sqrt{3}}\left(t-\frac{1}{2}\right)^2=\frac{\pi n}{|j|},
\end{equation}
which leads to the expression Eq.~(\ref{energies-beta=1/2}) in the main text.

\section{Limit $E=0$  from Eq.~(\ref{int-general})}
\label{sec:E=Delta=0}

Here we show that the Bohr-Sommerfeld quantization condition Eq.~(\ref{BS-condition})
with the LHS given by Eq.~(\ref{int-general}) in the absence of an electric field produces the spectrum Eq.~(\ref{LLspectrum-magfield}) with $\Delta=0$.
For $\beta=0$ the first term of Eq.~(\ref{int-general}) disappears,
\begin{equation}
\label{int-general-beta=0}
I(\epsilon,j,\beta=0)=\hbar
\left[(\epsilon^2-j)J_1-2j^2J_{-1}\right],
\end{equation}
and as mentioned above Eq.~(\ref{roots>0}), the arguments of $J_{-1}$ and $J_1$
have opposite values $d=-a$ and $c=-b$.
Then Eqs.~(\ref{J1}) and (\ref{J-1}) are simplified to the form
\begin{subequations}
\label{Js-beta=0}
\begin{align}
\label{J1-beta=0}
J_{1} =& \frac{2}{a+b}\left[2b \Pi\left(k,k\right)
- b K(k)\right], \\
\label{J-1-beta=0}
J_{-1}= &\frac{2}{a+b}\left[\frac{2}{b}\Pi\left(-k,k\right)
-\frac{1}{b} K(k)\right],
\end{align}
\end{subequations}
with $k=(a-b)/(a+b)$.

For definiteness, we choose  the case $\epsilon>0,j>0$ corresponding
to the first line of Eq.~(\ref{roots>0}) that gives
\begin{equation}
\begin{split}
a=\epsilon+\sqrt{\epsilon^2-2j}, &\qquad b=\epsilon-\sqrt{\epsilon^2-2j},\\ k=&\frac{\sqrt{\epsilon^2-2j}}{\epsilon}.
\end{split}
\end{equation}
Rewriting the relation Eq.~(19.6.2) from Ref.~\cite{NIST-book} as
\begin{equation}
\Pi(\pm k,k)= \frac{\pi}{4(1\mp k)}+\frac{1}{2}K(k),
\end{equation}
one can verify that Eq.~(\ref{int-general-beta=0}) reduces to
\begin{equation}
\label{int-general-beta=0-fin}
I(\epsilon,j,\beta=0)=\hbar \frac{\pi}{2}
(\epsilon^2-2j),
\end{equation}
which agrees with  Eq.~(\ref{int-E=0}) for $\delta=0$
and $j>0$.
This completes the proof that one can recover the spectrum
(\ref{LLspectrum-magfield}) with $\Delta=0$ from the
quantization condition Eq.~(\ref{BS-condition}) with the LHS given by
Eq.~(\ref{int-general}) by setting  $\beta=0$.

\section{Shooting method}
\label{sec:shooting}

Equation~(\ref{matrix-M}) for the spinor components
$f(\rho)$ and $g(\rho)$ defined by Eq.~(\ref{angular-spinor})
has the form:
\begin{equation}
\label{stsys}
\begin{split}
\frac{df(\rho)}{d\rho} = \left(\frac{j}{\rho}+\frac{\rho}{2}\right)f(\rho) + \left(\beta\rho-\varepsilon -\delta\right)g(\rho),\\
-\frac{dg(\rho)}{d\rho} = \left(\frac{j}{\rho} + \frac{\rho}{2}\right)g(\rho) + \left(\beta\rho-\varepsilon + \delta\right)f(\rho).
\end{split}
\end{equation}
One can obtain from Eqs.~(\ref{stsys}) the derivative
$df/dg$.  It can be integrated up to a constant which is taken to
be equal to zero, because the subsequent equation should have a trivial solution $f=g=0$ as $\rho\xrightarrow[]{}\infty$:
\begin{equation}
    f^2\left(\beta\rho - \varepsilon + \delta\right)+2gf\left(\frac{j}{\rho} + \frac{\rho}{2}\right)+g^2\left(\beta\rho - \varepsilon - \delta\right)=0.
\end{equation}
Thus, the ratio $f(\rho)/g(\rho)$ can be obtained. Using it,
we can guess the initial value for one function and then calculate
the initial value for the other one. The boundary values at $\rho \xrightarrow[]{}\infty$ are expected to be zero, since functions
$f(\rho), g(\rho)$ have to be square integrable.

Now, when we know the initial and boundary conditions, we perform an integration process on the system Eq.~(\ref{stsys}) setting the numeric value of $\varepsilon$ from an interval of interest. Such a method is called a shooting method, for more details see Ref.~\cite{PresShooting}. To do the integration we use the Runge-Kutta sixth-order method. The boundaries for integration are $\rho\in[0.001, 100]$,
where the beginning of the interval replaces zero (which is a point of singularity
in the system) and the end replaces the infinity
(in this case, $r = l \rho$ is comparable to the length of the chain
used in the diagonalization method).

Shooting with energies from an interval of interest and performing a numerical integration from some initial values one can obtain a dependence of $f(\rho\xrightarrow[]{}\infty)$ [or $g(\rho\xrightarrow[]{}\infty)$] on
the energy $\varepsilon$. Such a function goes through zero every time the energy value is guessed correctly. This allows us to use a bisection method
on the mentioned dependencies to calculate the energies.


\begin{thebibliography}{99}
\bibitem{foot1} Note that SI units are used, for example, in
Ref.~\cite{Arjona20017PRB}. Accordigly, Lorenz transformations
for the fields and other equations in SI units are recovered
by replacing $H \to c B$.
\end{thebibliography}
\end{document}